# A k·p treatment of edge states in narrow 2D topological insulators, with standard boundary conditions for the wave function and its derivative

P.C. Klipstein

Semiconductor Devices, P.O. Box 2250, Haifa 31021, Israel

Abstract

For 2D topological insulators with strong electron-hole hybridization, such as HgTe/CdTe quantum wells, the widely used $4 \times 4$ $\mathbf{k} \cdot \mathbf{p}$ Hamiltonian based on the first electron and heavy hole sub-bands yields an equal number of physical and spurious solutions, for both the bulk states and the edge states. For symmetric bands and zero wave vector parallel to the sample edge, the mid-gap bulk solutions are identical to the edge solutions. In all cases, the physical edge solution is exponentially localized to the boundary and has been shown previously to satisfy standard boundary conditions for the wave function and its derivative, even in the limit of an infinite wall potential. The same treatment is now extended to the case of narrow sample widths, where for each spin direction, a gap appears in the edge state dispersions. For widths greater than 200 nm, this gap is less than half of the value reported for open boundary conditions, which are called into question because they include a spurious wave function component. The gap in the edge state dispersions is also calculated for weakly hybridized quantum wells such as InAs/GaSb/AlSb. In contrast to the strongly hybridized case, the edge states at the zone center only have pure exponential character when the bands are symmetric and when the sample has certain characteristic width values.

## I. INTRODUCTION

Semiconductor quantum wells (QWs) containing elements with a large atomic number have both a narrow band gap and a large spin-orbit splitting. For layer thicknesses greater than a critical value it is often possible to reverse the ordering of the *s*- and *p*-symmetry confined states so that the material undergoes a transition to a two dimensional topological insulator phase [1]. Since the spin-orbit splitting is also large, the topological phase can develop a significant band gap remote from other states, whereby the bulk material is an insulator at low temperatures, while the edges conduct massless Dirac Fermions whose spin is locked to their direction of motion. Two prime examples are HgTe/CdTe [2] and InAs/GaSb/AlSb [3]. Their band profiles with an inversion of the *s*- and *p*- states are shown schematically in Figure 1. These materials exhibit properties such as the quantum spin Hall effect [4, 5] where opposite spins are separated to different edges of a bar shaped sample. The edge conductance is also quantized in units of $e^2/h$ per edge.

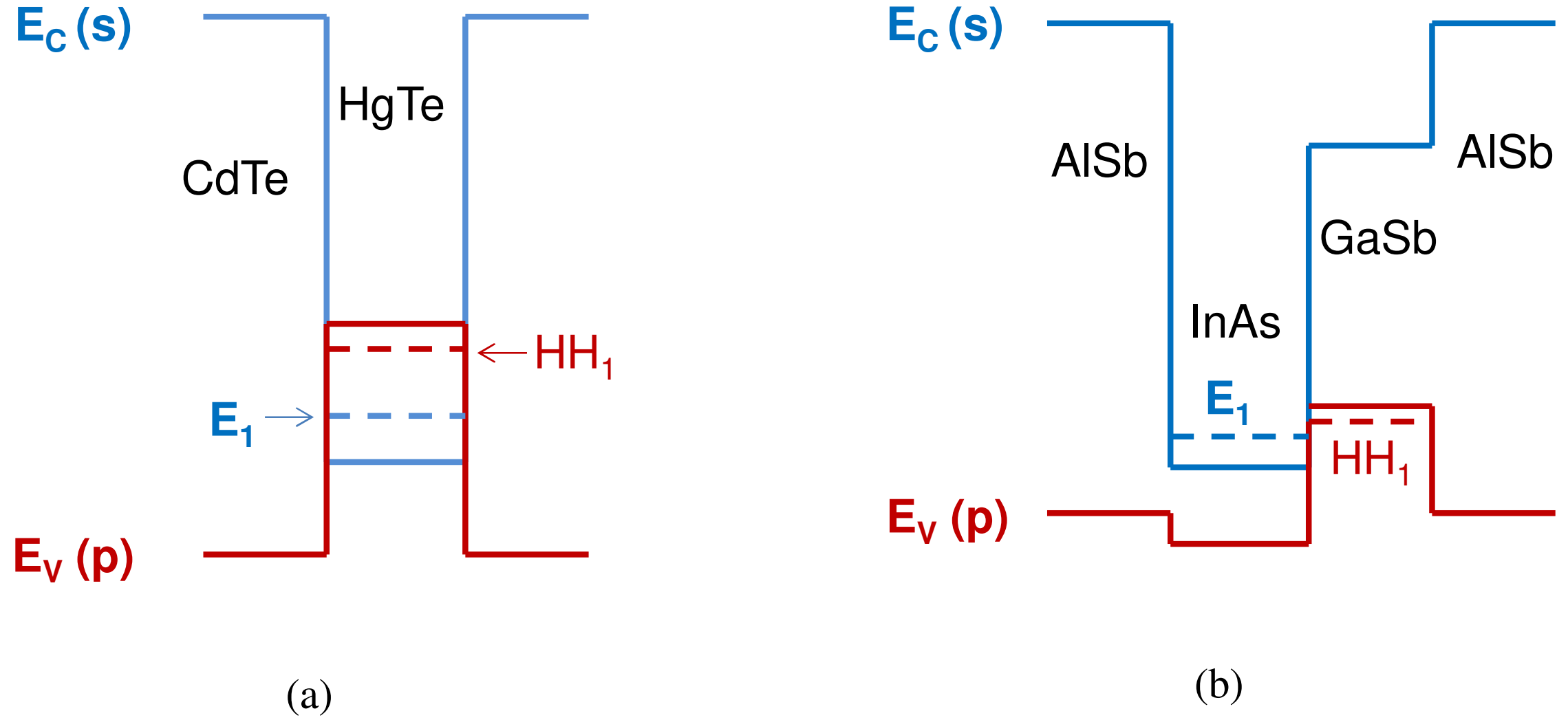

(a) (b)

**Figure 1.** Band profiles in the growth direction for a 2D topological insulator QW based on (a) HgTe/CdTe or (b) InAs/GaSb/AlSb. Note the inverted ordering of the electron, $E_1$, and heavy hole, $HH_1$, ground states, characteristic of the topological phase.

The growth of II-VI $Hg_{1-x}Cd_xTe$/CdTe QWs is very challenging, due to their low growth temperatures and material quality issues [6]. On the other hand III-V InAs/GaSb/AlSb QWs can be grown with high quality, and superlattices based on these materials are currently being developed both academically and commercially for various applications including as photodetector and laser materials [7, 8]. Therefore, the III-V material is likely to be more accessible to a wider range of institutions for the study of topological phenomena. Moreover, recent reports have shown that in spite of their smaller topological insulator band gaps, III-V QWs show remarkably robust edge states with properties not inferior to those of II-VI QWs. For example, a very precise quantized conductance has been reported in InAs/GaSb/AlSb QWs that persists up to about 5 K [9], and the edge state coherence length appears to be several microns [10].

A significant difference between the II-VI and III-V QWs is the strength of the electron-hole hybridization which, due to its type II band alignment, is about an order of magnitude weaker in the III-V system. As shown previously using a Hamiltonian based on $\mathbf{k}\cdot\mathbf{p}$ theory, this results in wave functions whose different character may perhaps account for some of the differences observed in the physical behavior of the two topological insulator materials [11,12]. In the II-VI material the wave function decays monotonically with distance from the edge, while in the III-V material it decays faster and oscillates. The previous work treated a single edge of a QW with a semi-infinite sample width. In the present work the treatment is extended to narrow sample widths, where states from opposite edges begin to overlap. It was shown by Zhou et al., that for sufficiently narrow HgTe/CdTe samples, a gap opens up in the edge state dispersions [13]. Their analysis used open boundary conditions (OBCs) which impose zero amplitude on the wave function at the sample edge. As argued previously and in this work, such an approach combines spurious "wing" solutions with real physical solutions in order to satisfy the zero amplitude condition, and this can lead to unphysical results [14]. For example, in wide samples, minor variations of the electron-hole dispersion asymmetry can cause the edge state velocity to vanish, or the edge state Dirac point to shift from one bulk band edge to the other, even when the dispersion asymmetry terms are negligibly small.

| *Parameter Set* | *A* (eVÅ) | *B* (eVÅ$^2$) | *D* (eVÅ$^2$) | *M* (eV) | *Ref.* |
|---|---|---|---|---|---|
| 1 | 3.65 | 68.6 | 51.2 | -0.010 | [13] |
| 2 | 3.86 | 50 | 70 | -0.010 | [11] |
| 3 | 0.46 | 56 | 0 | -0.010 | [11] |

**TABLE I.** Parameters used in the calculations.

Due to the failure of OBCs, the problem of interacting edge states in narrow width samples is readdressed in this work using standard boundary conditions (SBCs) for the physical wave function and its derivative, which are obtained by integrating the Hamiltonian across the edge region. It was shown previously that SBCs can be satisfied by exponentially localized edge states, even when the potential of the sample wall is infinite [11, 12]. For narrow samples with strong electron-hole hybridization, the gap in the edge state dispersion is significantly smaller than that found using OBCs. This gap is also calculated for narrow weakly hybridized systems, using parameters typical of InAs/GaSb/AlSb quantum wells. In contrast to the strongly hybridized case, the edge states in weakly hybridized systems with symmetric band dispersions only decay exponentially for certain edge state wave vectors close to the zone center and certain equally spaced sample widths. When the dispersions are asymmetric there are no exponential solutions, i.e. the decay is non-exponential, but in all cases exponential character is re-established close to the merging points of the edge states with the bulk band edges.

This paper is arranged as follows. In the next section a model Hamiltonian based on the lowest electron and heavy-hole sub-bands is introduced, which is similar to that used by many other workers. As discussed above, OBCs cannot be used, so the properties of the Hamiltonian must be considered in both the sample and wall regions. The issue of spurious solutions is discussed in the following section where a clear relationship is shown between spurious bulk and edge solutions of two band

Hamiltonians. SBCs are then derived as a natural alternative to OBCs. Their solutions are shown to be consistent with the topological Chern numbers of the sample and its boundaries. The remainder of the paper deals with a derivation of the SBC characteristic equation for samples of any width, and an analysis of its properties. A simple analytical formula is obtained for the edge state band gap in narrow samples, and in the final section of the paper, possible comparisons with experiment are suggested. It should be emphasized that the parameters proposed for the model Hamiltonian (sets 2 and 3 in Table I) were derived in Ref. [11] using an 8 × 8 **k · p** calculation that has stood up well to a comparison with experiment [15, 16], so the results of the present study should be reasonably representative. Both analytical and numerical approaches have shown that structural asymmetry terms have only a minor influence on the edge state energies, and they are not considered in this work [12, 17].

## II. MODEL HAMILTONIAN

A 4×4 block-diagonal **k · p** Hamiltonian for a 2D topological insulator QW sample is used here, which is similar to that described in several previous works based on **k · p** and other approaches [2, 4, 11, 13]. Considering only spin up states, a 2×2 wave equation in a basis of the first confined electron and heavy-hole states can be taken from its upper block as follows:

$$\begin{bmatrix} M+\mathbf{k}\cdot B_{+}\mathbf{k} & A\left(k_x+ik_y\right) \\ A\left(k_x-ik_y\right) & -M-\mathbf{k}\cdot B_{-}\mathbf{k} \end{bmatrix}\begin{bmatrix} \psi_{1\uparrow} \\ \psi_{2\uparrow} \end{bmatrix}=E\begin{bmatrix} \psi_{1\uparrow} \\ \psi_{2\uparrow} \end{bmatrix} \tag{1}$$

The equivalent 2×2 wave equation for spin down states is taken from the lower block, although its solutions can be obtained from Eq. (1) by time reversal. In Eq. (1), $A$ is the electron-hole hybridization parameter and $\mathbf{k}=\left(k_x,k_y\right)$ is the wave vector in the plane of the QW. The parameters, $B_{\pm} = B \pm D$, are proportional to inverse band edge effective masses which are determined by second order interactions with remote states. In Eq. (1), $B > 0$, $D > 0$, and in the topological phase the band order is inverted, so $M < 0$, where $2|M|$ is the zone center QW band gap. In the wall regions at the boundaries of the sample $B = B_0$, $D = D_0$, $A = A_0$ and $M = M_0 > 0$. As in previous works, Eq. (1) will be solved for the edge states in the limit $B_0 = D_0 = 0$, when the wall obeys a relativistic Hamiltonian for a free particle with energies: $E = \pm\sqrt{p^2c^2 + m_0^2c^4}$, where $p^2c^2 = A_0^2\left(k_x^2 + k_y^2\right)$ and $m_0c^2 = M_0$. For $M_0 \to \infty$, the wall provides an infinite confining potential for both electrons and holes [11].

Previously, Eq. (1) was solved with $A_0 = A$ [11, 12]. This model implicitly assumes that the crystal periodic basis states are the same in both the sample and the wall, and all other crystal periodic basis states are extremely remote [18, 19]. It avoids having to know the ordering of the off-diagonal electron-hole hybridization operators, since there are no "interface" contributions at the sample edges due to operator ordering. In principal any residual δ-function like edge terms of the type discussed in Refs. [18] and [19] can be included as a perturbation, as discussed at the end of section 2 of Ref. [12]. However, since the edge states penetrate the sample over much greater distances than the bulk lattice constant, it will be assumed that their amplitude at the edge is small enough that such edge terms can be ignored.

Since the wall potential, $M_0$, is very different from the $M$-value in the sample, it is probably a better assumption to suppose that the $A$-parameters in the sample and wall are also different. Realistic wall parameters cannot be derived through perturbation theory, so a more phenomenological approach must be taken. In the following treatment, most results are presented for the simpler case of $A_0 = A$, but the effect of different $A$-values in the sample and wall regions is also investigated. In all cases, interface contributions from the edges are neglected. Fortunately, even when $A_0 = \hbar c >> A$ (corresponding to a totally free particle), the results for strongly hybridized HgTe/CdTe QWs are virtually the same as for $A_0$ = $A$. For weakly hybridized systems with symmetric bands ($D$ = 0), the results are totally independent of $A_0$.

## III. BOUNDARY CONDITIONS

Eq. (1) with $D$ = 0 and $k_x$ = 0 corresponds directly with the 2×2 **k · p** Hamiltonian of White and Sham [20] who showed that for small $B < A^2/4|M|$, it produces two pairs of bulk evanescent states in the band gap energy range, $|M| > E > -|M|$, one of which they termed the "wing" bands and the other the "middle" bands. The "middle" bands are physical evanescent states with small imaginary wave vector, $iq_m(E) \approx \pm i\sqrt{(M-E)(M+E)}/A$, that correspond, for example, to a physically decaying wave function when the material is used as a barrier in a tunnel structure. The "wing" bands, on the other hand, do not describe any real physical states. Their imaginary wave vectors are $iq_w(E) \to \pm i\frac{A}{B}$ and can become very large, with a magnitude that can even go beyond the first Brillouin zone boundary. They are a mathematical necessity for matching the boundary conditions on the wave-function and its derivative, but do not exist in a more complete solution involving a larger set of basis states [21].

Topological arguments show that for $M_0 \to +\infty$, $M < 0$, $D = 0$, Eq. (1) has a spin polarized edge solution travelling in the $x$-direction with energy $-Ak_x$ (the equivalent spin down solution is $+Ak_x$) [4, 22]. When $B$ is non-zero, there are actually two degenerate edge solutions of Eq. (1) which have wave functions at zero energy and wave vector proportional to $\begin{bmatrix} 1 \\ -1 \end{bmatrix} e^{-q_m(0)y}$ and $\begin{bmatrix} 1 \\ -1 \end{bmatrix} e^{-q_w(0)y}$. These evanescent edge solutions of Eq. (1) at $k_x$ = 0 are no different from the bulk case, discussed above. A common procedure is to apply OBCs for an edge at $y$ = 0, yielding a single wave function at $k_x$ = 0 of the form $\begin{bmatrix} 1 \\ -1 \end{bmatrix} \left( e^{-q_m(0)y} - e^{-q_w(0)y} \right)$ which has zero amplitude at the edge [13, 23, 24]. Unfortunately, this procedure is not valid since the physical "middle" solution is combined with the unphysical "wing" solution that would not exist in a more complete treatment, as discussed above. Moreover, for non-zero *D*, there is still a middle solution, and a spurious evanescent "wing" solution which becomes a spurious oscillatory solution in materials with $D > B$, as pointed out by Schuurmans and t'Hooft [21]. There is then no edge solution that satisfies OBCs, yet Schuurmans and t'Hooft argued that precisely this condition occurs in many bulk materials (e.g. GaAs and AlAs) [21], and the author previously showed that $D > B$ also applies to HgTe/CdTe QWs [11]. Finally, for $B = D = 0$, there are no wing solutions and only one exponential solution with an energy independent decay parameter equal to $q_m(0)$ [12, 14, 22]. In this case it is impossible to satisfy OBCs so they have no solution, contrary to the topological argument at the end of this section where a single edge state is predicted.

Because OBCs take the spurious solutions completely seriously, this can lead to non-physical consequences which have been discussed by the author in Ref. [12]. For example, they predict an edge state dispersion which varies as $E = -M\frac{D}{B} \pm Ak\sqrt{1-\left(\frac{D}{B}\right)^2}$ [13], and in strongly hybridized systems this yields a particle velocity which is zero when $D = B$, even when these remote interaction parameters are vanishingly small and the remote states should have no influence. The "spurious" or "wing" solution must therefore be rejected, and an alternative treatment is required for a wave function with a finite amplitude at the edge. This wave function must satisfy boundary conditions which take the wall region into account. These boundary conditions, termed SBCs in the previous section, are derived for the wave function and its derivative by integrating Eq. (1) across the wall region. For a wall Hamiltonian as defined in section II, and an edge at $y = 0$, the SBCs are:

$$\psi_{i\uparrow}\Big|_{-\varepsilon} = \psi_{i\uparrow}\Big|_{+\varepsilon} \qquad \text{(a)}$$

$$B_{0\pm}\frac{\partial \psi_{i\uparrow}}{\partial y}\Bigg|_{-\varepsilon} = B_{\pm}\frac{\partial \psi_{i\uparrow}}{\partial y}\Bigg|_{+\varepsilon} \qquad \text{(b)} \qquad (2)$$

with positive $\varepsilon \to 0$, and where the positive subscript of the $B$-parameters in Eq. (2) (b) applies to $i$ = 1 and the negative to $i$ = 2. Based on this SBC approach, exponentially decaying solutions have been derived for the edge of a semi-infinite sample in two earlier works in the limit of $M_0 \to +\infty$, corresponding to an infinite confining potential beyond the sample edge [11,12]. The problem of zero edge state velocity for $D = B$ is removed in the SBC treatment, because the dispersion essentially has the form: $Dk^2 \pm Ak$, which simply becomes more linear as the remote states become more remote ($D$, $B \to 0$). The two approaches are compared in Figure 2, for a semi-infinite sample based on parameters from Ref. [13]. This figure demonstrates a rapid and unphysical variation of the edge state velocity in the OBC picture when the $D$-parameter is slightly increased, from $D < B$ to $D > B$. In contrast to the OBC solutions which only exist for $D < B$, SBC solutions exist for all $D$-values. They merge smoothly with the bulk states which hardly change as $D$ is increased. Such strong variation of the OBC edge states due to weak remote interactions is quite unphysical. Even if the remote interactions are ten or a hundred times weaker, the picture remains virtually the same.

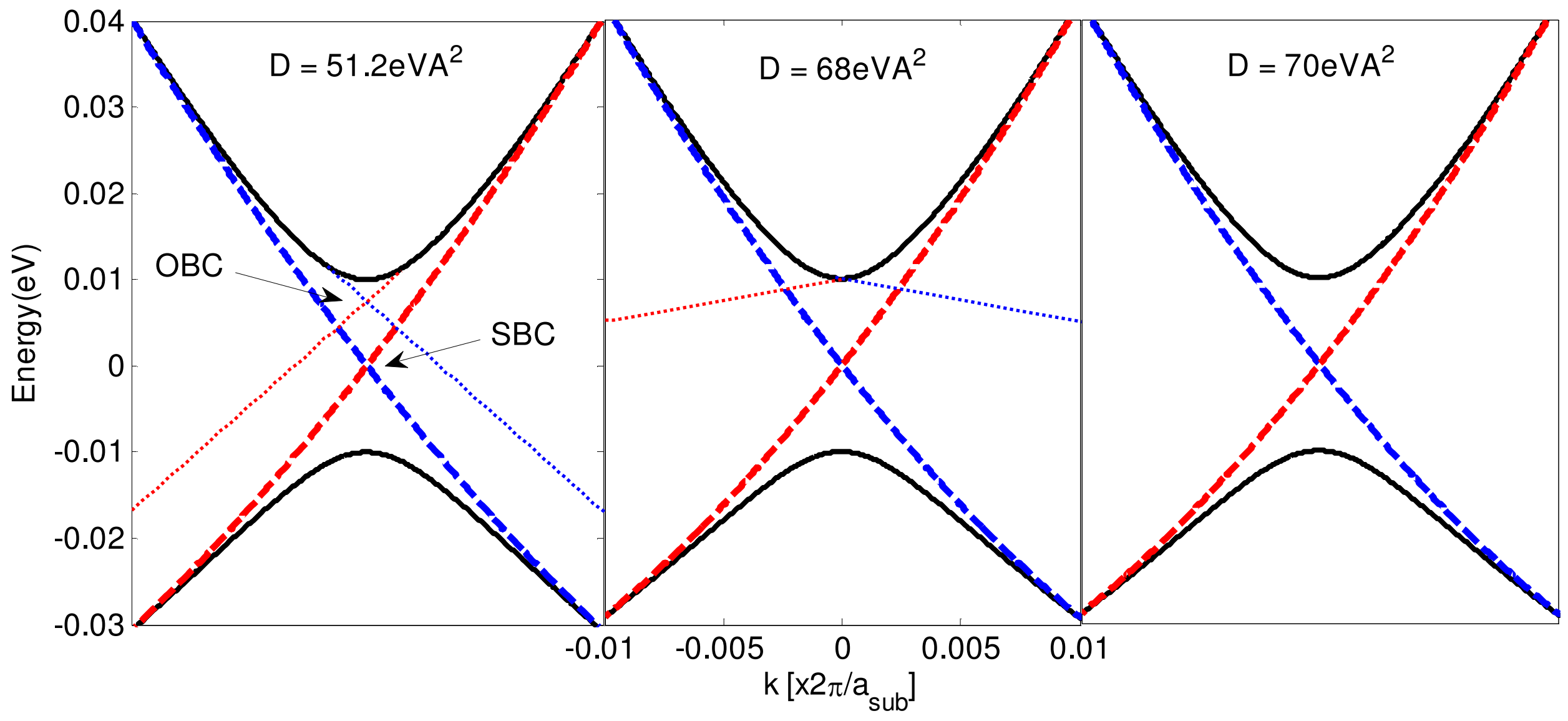

**Figure 2.** Dispersions of bulk states (black lines) and edge states for a HgTe/CdTe QW sample of semi-infinite width calculated using OBCs (dotted lines) and SBCs (dashed lines), based on parameter set 1 in Table I (spin up is blue and spin down is red). In the middle and right hand panels, the value of the $D$ parameter has been increased from 51.2, to 68 and 70 eV Å$^2$, respectively. In the right hand panel there is no OBC solution, while in the middle one the OBC edge state velocity is very small. The cubic lattice parameter of the substrate is $a_{sub}$ = 6.453Å.

In the remainder of this work, the SBC treatment is extended to edge states in narrow samples of width, $L_S = 2L$ ($-L < y < L$), so that a non-zero edge state amplitude exists on both sample edges. Before deriving their characteristic equation, this section concludes with a discussion of some topological aspects of the SBC treatment. For simplicity, it is assumed that the band structure is symmetric ( $D_0 = D = 0$ ) and only a single edge is considered. For consistency with the earlier part of this section, it is also assumed that $B$ is small. Exponentially confined edge states, with real decay parameters σ and $\sigma_0$, in the sample and wall respectively, exist for negative $B_0$, since Eq. (2) (b) yields $B_0 = \frac{\sigma}{\sigma_0} B$, and the decay parameters have opposite signs [11]. A wall with $M_0 > 0$, $B_0 < 0$ and $A_0 > 0$ is a topological phase characterized by a Chern number of −1, while the sample with $M < 0$, $B > 0$ and $A > 0$ is a topological phase characterized by a Chern number of +1. The difference in Chern numbers is two, so there should be two edge state solutions for any negative value of $B_0$, guaranteed to propagate in the same direction along the boundary [25]. The envelope functions of the two edge states with a given wave vector, $k_x$, are depicted schematically along the $y$-direction in Figure 3 (a). In Ref. [11], two solutions were found analytically using the boundary conditions in Eq. (2). These solutions were exponential only for a specific value of $B_0$ which gave energies of $E = -Ak_x$ for spin up and $E = +Ak_x$ for spin down (non-exponential solutions will exist for other values of $B_0$ with different dispersion energies). It was shown that when $M_0 \to \infty$, the decay parameter in the wall goes to infinity ($\sigma_0 \to \infty$) and this value of $B_0$ tends to zero, under which circumstances the wall, termed wall (1) in Figure 3 (a), reasonably describes a vacuum based on the massive Dirac Hamiltonian. Since $E = 0$ when $k_x = 0$, one of the two solutions for the infinite wall can be eliminated as a spurious wing solution, as discussed for the identical case earlier in this section, because it would not exist for a more realistic sample Hamiltonian that is periodic in $k$-space and derived with a larger basis. This solution is depicted by the dashed line in Figure 3 (a). An alternative argument that allows the non-physical edge state to be eliminated is to take an intermediate wall layer, termed wall (2) in Figure 3 (b) with a positive $B$-parameter, $B_0' > 0$, and insert it in-between wall (1) and the sample of Figure 3 (a). In the limit $M_0 \to \infty$, wall (2) is the most general representation of a topologically trivial vacuum, with a Chern number of zero and a Chern number difference of one with the sample, based only on a change in sign of the $M$-parameter. The sample edge now only supports a single edge state, namely the one that decays more slowly in the sample, a result that is quite robust since it only depends on the band ordering and not on any changes of the dispersions. The second edge state moves to the left hand interface of wall (2) where the Chern number again changes by one, this time due to a change in sign of the quadratic *B*-terms. Wall (2) can be expanded to a semi-infinite width, removing the second edge state and leaving a single, stable edge state at the sample boundary. In the limit $B_0' \to 0$, this edge state will evolve into the same physical edge solution described above when $B_0 \to 0$ and depicted by the solid line in Figure 3 (a).

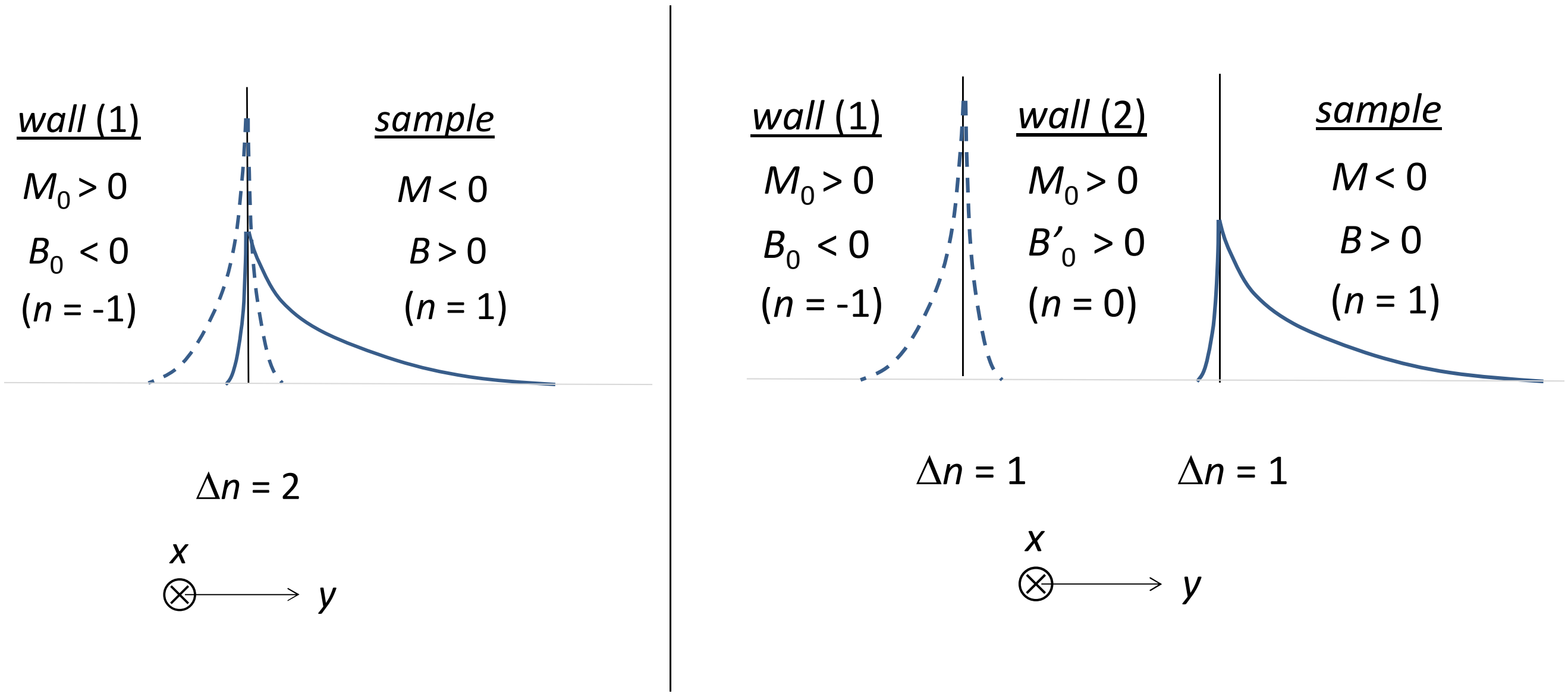

(a) (b)

**Figure 3.** (a) Sample-wall interface parallel to the $x$-axis with a Chern number difference of $\Delta n = 2$, schematically depicting the amplitude of two edge state solutions with the same spin propagating in the same $x$-direction (solid and dashed lines). Wall (1) has vacuum like properties when $B_0 = 0$ (b) Insertion of wall (2), a general vacuum-like intermediate layer with zero Chern number that separates the edge states to its interfaces with the other two layers (the $A$-parameters for all three layers are taken to be positive).

Finally, the case where the *B*-parameters in both the sample and wall are zero has been discussed by Tkachov and Hankiewicz [22] who found a single exponentially decaying solution of Eq. (1) using a condition of zero current perpendicular to the boundary. In the SBC treatment, boundary condition (2) (b) vanishes and Eq. (1) yields a single edge solution identical to that of Tkachov and Hankiewicz, which also corresponds with the physical edge solution discussed above when evaluated in the same limit [11]. Although each phase in this case is topologically trivial, the change in Chern number across the boundary is one, consistent with the existence of a single topological edge state.

## IV. CHARACTERISTIC EQUATION FOR EDGE STATES IN NARROW SAMPLES

In this section, edge state solutions are sought which are exponentially localized to the sample boundaries at $y = \pm L$, when the confining potential of the walls is very large ($M_0 \to \infty$). As for the semi-infinite case discussed above, there are only exponential solutions when the dispersion parameters in the walls tend to zero, namely $B_{0L}$, $D_{0L} \to 0$ for the left hand wall and $B_{0R}$, $D_{0R} \to 0$ for the right, and the vacuum surrounding the sample is then described by the massive Dirac Hamiltonian.

The wave function for a narrow sample can be defined as follows:

$$\begin{bmatrix}\psi_{1\uparrow}\\ \psi_{2\uparrow}\end{bmatrix}=\begin{bmatrix}A_1\cosh\sigma y+B_1\sinh\sigma y\\ A_2\cosh\sigma y+B_2\sinh\sigma y\end{bmatrix}e^{ikx}=\overline{\psi}_{\uparrow,\sigma}(y)e^{ikx} \quad (3)$$

where $k = k_x$ is the edge state wave vector, and $\sigma$ is the exponential decay parameter in the sample. The constants $A_1$, $B_1$, $A_2$, $B_2$ are determined by the boundary conditions. Substituting Eq. (3) into Eq. (1), an equation of the form: $\overline{\overline{D}}\overline{A}=0$ is obtained where $\overline{A}$ is a column vector with components: $A_1$, $B_1$, $A_2$, $B_2$. Solving $|D|=0$ yields two solutions for the decay parameter, $\sigma$:

$$\sigma_\pm=\sqrt{k^2+F\pm\sqrt{F^2-G}} \quad (4)$$

where $F=\frac{A^2+2(MB+ED)}{2B_+B_-}$ and $G=\frac{M^2-E^2}{B_+B_-}$. This is identical to the formula derived previously for a single edge [11, 12, 26]. Decay parameters with a given energy are either both real, real and imaginary, or complex conjugates. Contrary to previous works [11, 12], where a prime was added when $D\neq 0$, i.e. $\sigma'_\pm$, the prime is dropped in the present work and the same symbol, $\sigma_\pm$, is used in all cases.

The wave functions in the wall region are simple exponentially decaying functions proportional to $\exp(-\sigma_{0L}[y+L])$ for the left hand wall and $\exp(\sigma_{0R}[y-L])$ for the right, where the decay parameters, $\sigma_{0L}$ and $\sigma_{0R}$ are both negative and tend to infinity when $M_0\rightarrow+\infty$. Applying the SBC boundary conditions for the wave function derivatives in Eq. (2)(b) yields the following relation for the coefficients:

$$\frac{A_1}{B_1}=\frac{A_2}{B_2}=-\frac{\sigma-(r_L\sigma_{0L})\tanh\sigma L}{(r_L\sigma_{0L})-\sigma\tanh\sigma L}=\frac{\sigma-(r_R\sigma_{0R})\tanh\sigma L}{(r_R\sigma_{0R})-\sigma\tanh\sigma L} \quad (5)$$

where $r_L = B_{0L}/B = D_{0L}/D$ and $r_R = B_{0R}/B = D_{0R}/D$. In the limit, $M_0 \to +\infty$, it is shown below that $r_L \to 0$ and $r_R \to 0$, while the products $r_L\sigma_{0L}$ and $r_R\sigma_{0R}$ tend to finite values.

Applying the SBC boundary condition for continuity of the wave-function in Eq. (2)(a), and substituting expressions for the wave function into the Hamiltonian equation for the wall, produces the following result in the limit $M_0 \to +\infty$:

$$\begin{bmatrix} \left(1-\frac{\sigma_{0L}^2 B_{0L+}}{M_0}\right)\cosh\sigma L & -\left(1-\frac{\sigma_{0L}^2 B_{0L+}}{M_0}\right)\sinh\sigma L & -A_0\left(\frac{\sigma_{0L}}{M_0}\right)\cosh\sigma L & A_0\left(\frac{\sigma_{0L}}{M_0}\right)\sinh\sigma L \\ -A_0\left(\frac{\sigma_{0R}}{M_0}\right)\cosh\sigma L & -A_0\left(\frac{\sigma_{0R}}{M_0}\right)\sinh\sigma L & -\left(1-\frac{\sigma_{0R}^2 B_{0R-}}{M_0}\right)\cosh\sigma L & -\left(1-\frac{\sigma_{0R}^2 B_{0R-}}{M_0}\right)\sinh\sigma L \\ A_0\left(\frac{\sigma_{0L}}{M_0}\right)\cosh\sigma L & -A_0\left(\frac{\sigma_{0L}}{M_0}\right)\sinh\sigma L & -\left(1-\frac{\sigma_{0L}^2 B_{0L-}}{M_0}\right)\cosh\sigma L & \left(1-\frac{\sigma_{0L}^2 B_{0L-}}{M_0}\right)\sinh\sigma L \\ \left(1-\frac{\sigma_{0R}^2 B_{0R+}}{M_0}\right)\cosh\sigma L & \left(1-\frac{\sigma_{0R}^2 B_{0R+}}{M_0}\right)\sinh\sigma L & A_0\left(\frac{\sigma_{0R}}{M_0}\right)\cosh\sigma L & A_0\left(\frac{\sigma_{0R}}{M_0}\right)\sinh\sigma L \end{bmatrix} \begin{bmatrix} A_1 \\ B_1 \\ A_2 \\ B_2 \end{bmatrix} = 0 \qquad (6)$$

The determinant of the square matrix must be zero, which occurs when one of the following two conditions is fulfilled:

$$A_0^2 = \left(\frac{M_0}{\sigma_{0L}} - \sigma_{0L}B_{0L-}\right)\left(\frac{M_0}{\sigma_{0L}} - \sigma_{0L}B_{0L+}\right) \qquad (7)(a)$$

$$A_0^2 = \left(\frac{M_0}{\sigma_{0R}} - \sigma_{0R}B_{0R-}\right)\left(\frac{M_0}{\sigma_{0R}} - \sigma_{0R}B_{0R+}\right) \qquad (7)(b)$$

These conditions can be rearranged to give:

$$\left(\frac{M_0}{\sigma_{0R}}\right) = \left(r_R\sigma_{0R}\right)B - \sqrt{\left(r_R\sigma_{0R}\right)^2 D^2 + A_0^2} \qquad (8)(a)$$

$$\left(\frac{M_0}{\sigma_{0L}}\right) = \left(r_L\sigma_{0L}\right)B - \sqrt{\left(r_L\sigma_{0L}\right)^2 D^2 + A_0^2} \qquad (8)(b)$$

Eq. (8) is very similar to the expression for a single edge that appears just before Eq. (9) in Ref. [11]. It shows that for an edge where the wave function has a finite amplitude, the wall decay parameter tends to infinity with $M_0$, while the products $r_L\sigma_{0L}$ or $r_R\sigma_{0R}$ are finite, corresponding to $B_{0L}, B_{0R} \to 0$ and $D_{0L}, D_{0R} \to 0$, as required. Using the results in Eq. (8), it is possible to solve Eq. (6) for the vector, $\overline{A}$, yielding relations for the amplitudes:

$$\frac{A_2}{B_1} = \tanh \sigma L \frac{\hat{\sigma}D + \sqrt{\hat{\sigma}^2 D^2 + A_0^2}}{A_0} \qquad \text{(a)}$$

$$\frac{A_1}{B_2} = -\tanh \sigma L \frac{\hat{\sigma}D - \sqrt{\hat{\sigma}^2 D^2 + A_0^2}}{A_0} \qquad (9) \quad \text{(b)}$$

$$\frac{A_1}{B_1} = \frac{A_2}{B_2} = \gamma_{\hat{\sigma}} \tanh \sigma L \qquad \text{(c)}$$

There are two solutions with either $\hat{\sigma} = r_R\sigma_{0R}$ and $\gamma_{\hat{\sigma}} = 1$ or $\hat{\sigma} = r_L\sigma_{0L}$ and $\gamma_{\hat{\sigma}} = -1$. Inspection of Eq. (3) shows that $\gamma_{\hat{\sigma}} = 1$ in Eq. (9)(c) corresponds to $\overline{\psi}_{\uparrow,\sigma}(-L) = 0$, and $\gamma_{\hat{\sigma}} = -1$, to $\overline{\psi}_{\uparrow,\sigma}(+L) = 0$. Hence Eq. (9) describes solutions with a finite amplitude on the right hand edge and zero amplitude on the left, or vice versa. The term, $\hat{\sigma}$, contains the wall decay parameter for the edge on which the wave function is localized. The value of $\hat{\sigma}$ is obtained by inserting Eq. (9)(c) into Eq. (5), yielding:

$$\hat{\sigma} = \sigma \left[ \frac{1 + \tanh^2 \sigma L}{2 \tanh \sigma L} \right] \qquad (10)$$

The decay parameter in the opposite wall is indeterminate, since the amplitude in that wall is zero.

The wave function in Eq. (3) can be written:

$$\begin{bmatrix} \psi_{1\uparrow} \\ \psi_{2\uparrow} \end{bmatrix} = B_2 \begin{bmatrix} \frac{A_1}{B_2} \left( \cosh \sigma y + \frac{1}{\left[ \frac{A_1}{B_1} \right]} \sinh \sigma y \right) \\ \left( \frac{A_2}{B_2} \cosh \sigma y + \sinh \sigma y \right) \end{bmatrix} e^{ikx} \qquad (11)$$

Substituting the expressions in Eq. (9) yields a pair of independent solutions for the two edges, which are combined linearly to form the general solution. This solution takes the form:

$$\begin{bmatrix} \Psi_{1\uparrow} \\ \Psi_{2\uparrow} \end{bmatrix} = C_1 \begin{bmatrix} Q_D\left(\tanh\sigma L\cdot\cosh\sigma y + C_2\cdot\sinh\sigma y\right) \\ C_2\tanh\left(\sigma L\right)\cdot\cosh\sigma y + \sinh\sigma y \end{bmatrix} e^{ikx} = \overline{\Psi}_{\uparrow,\sigma}\left(y\right)e^{ikx} \tag{12}$$

where

$$Q_D = \frac{\sqrt{\hat{\sigma}^2 D^2 + A_0^2} - \hat{\sigma}D}{A_0} \tag{13}$$

When the general solution is substituted back into the Hamiltonian in Eq. (1), the coefficient $C_2$, is obtained:

$$C_2 = \frac{-Ak}{\left[M + B_+\left\{k^2 - \sigma^2\right\} - E\right]Q_D + A\sigma\tanh\sigma L} \tag{14}$$

together with the characteristic equation:

$$F_C = \left[\sigma\tanh\sigma L + \frac{\left(M + B_-\left\{k^2 - \sigma^2\right\} + E\right)}{AQ_D}\right]\left[\frac{\sigma}{\tanh\sigma L} + \frac{M + B_-\left\{k^2 - \sigma^2\right\} + E}{AQ_D}\right] - k^2 = 0 \tag{15}$$

Note that the characteristic equation is invariant to the transformation: $D \to -D$ and $E \to -E$. In the limit $L \to \infty$, Eq.(15) reverts to the result for a single edge, namely:

$$F_E = \left[\sigma + \frac{\left(M + B_-\left\{k^2 - \sigma^2\right\} + E\right)}{AQ_D}\right] \pm k = 0 \tag{16}$$

where the positive sign is for a left hand edge and the negative is for a right. This result agrees with Eq. (9) for a left hand edge in Ref. [11].

For a symmetric band structure ($D$ = 0) Eq. (15) can be rearranged to provide an analytical result for the edge state band gap by making the following substitutions:

$$M = -\frac{A\sigma}{2}\left[\frac{1}{\tanh\sigma L} + \tanh\sigma L\right] + B\left(\sigma^2 - k^2\right) \quad \text{(a)}$$

$$\Delta E = A\sigma\left[\frac{1}{\tanh\sigma L} - \tanh\sigma L\right] = \frac{2A\sigma}{\sinh\left(2\sigma L\right)} \quad \text{(b)} \qquad (17)$$

leading to a dispersion equation:

$$E = \pm\sqrt{A^2k^2 + \left(\tfrac{\Delta E}{2}\right)^2} \qquad (18)$$

where $\Delta E$ is the edge state band gap. In the wide sample limit, when $L \to \infty$, Eq. (17)(a) yields decay parameters, $\sigma_{\pm} = \frac{A}{2B} \pm \sqrt{\frac{A^2}{4B^2} - \frac{|M|}{B} + k^2}$, in agreement with the result derived previously for a semi-infinite sample [11].

## V. RESULTS AND DISCUSSION

Using the relations in Eqs. (12)–(15), some examples of edge state dispersions and wave functions are calculated in this section, for narrow samples based on HgTe/CdTe and InAs/GaSb/AlSb QWs. A number of different parameter sets are used to allow comparison with earlier works by the author and others. The different parameter sets are listed in Table I, together with references to the works from which they are taken. As mentioned above, $A_0 = A$, unless stated otherwise.

Figure 4 shows the dispersion of the spin up edge states in a HgTe/CdTe sample of width 100 nm, based on Eq. (15) and parameter set (2). This set was derived in Ref. [11] by fitting the bulk band edge dispersion calculated from Eq. (1) to the equivalent dispersion determined with an 8 band **k·p** model. Since $D > B$, there is only one evanescent edge state solution, corresponding to $\sigma_+$ in Eq. (4). As discussed above, $\sigma_-$ is imaginary and yields a spurious oscillatory solution. The edge state energies for decay parameter, $\sigma_+$, and very large widths were calculated using Eq. (16) and presented in Fig. 2 (b) of Ref. [11]. They obey dispersions which are essentially of the form: $Dk^2 - Ak$ for the left hand edge and $Dk^2 + Ak$ for the right (where $\sigma_+ \to 0$ at the merging points with the bulk band edges), and are shown in Figure 4 as dashed lines. The small width solutions tend to these values with increasing edge state wave vector. At the zone center, however, an energy gap opens up of 2.9 meV. The dependence of the gap on sample width will be discussed below.

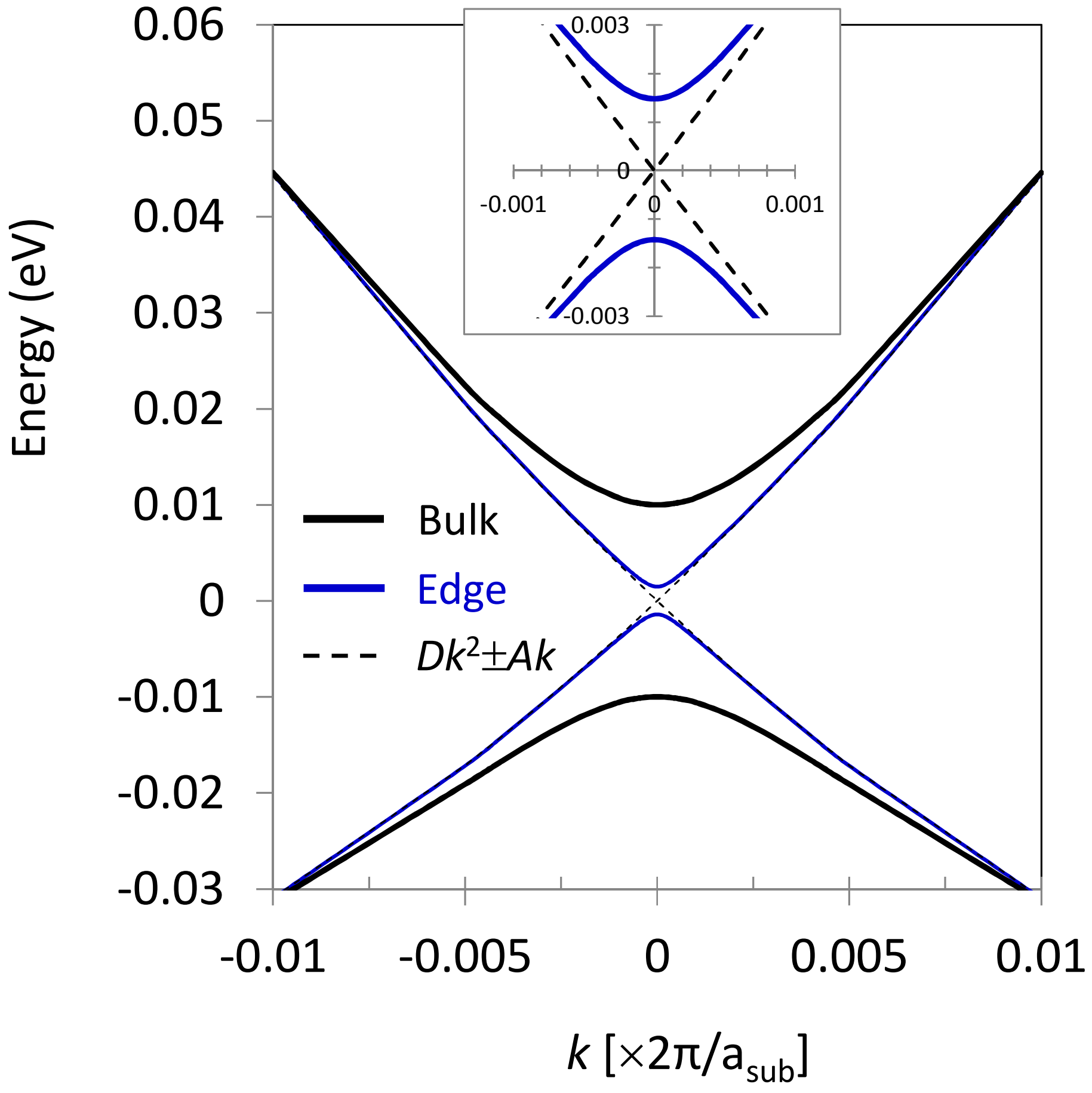

**Figure 4.** The solid blue lines show the spin up edge dispersions for a HgTe/CdTe topological insulator QW with a sample width of $L_S$ = 100 nm, calculated with parameter set (2). The bold black lines are the bulk dispersions. The dashed black lines show dispersions $D\,k^2 \pm A\,k$ which correspond to a very wide sample. The inset shows the zone center region on an expanded scale. The cubic lattice parameter of the substrate is $a_{sub}$ = 6.453Å.

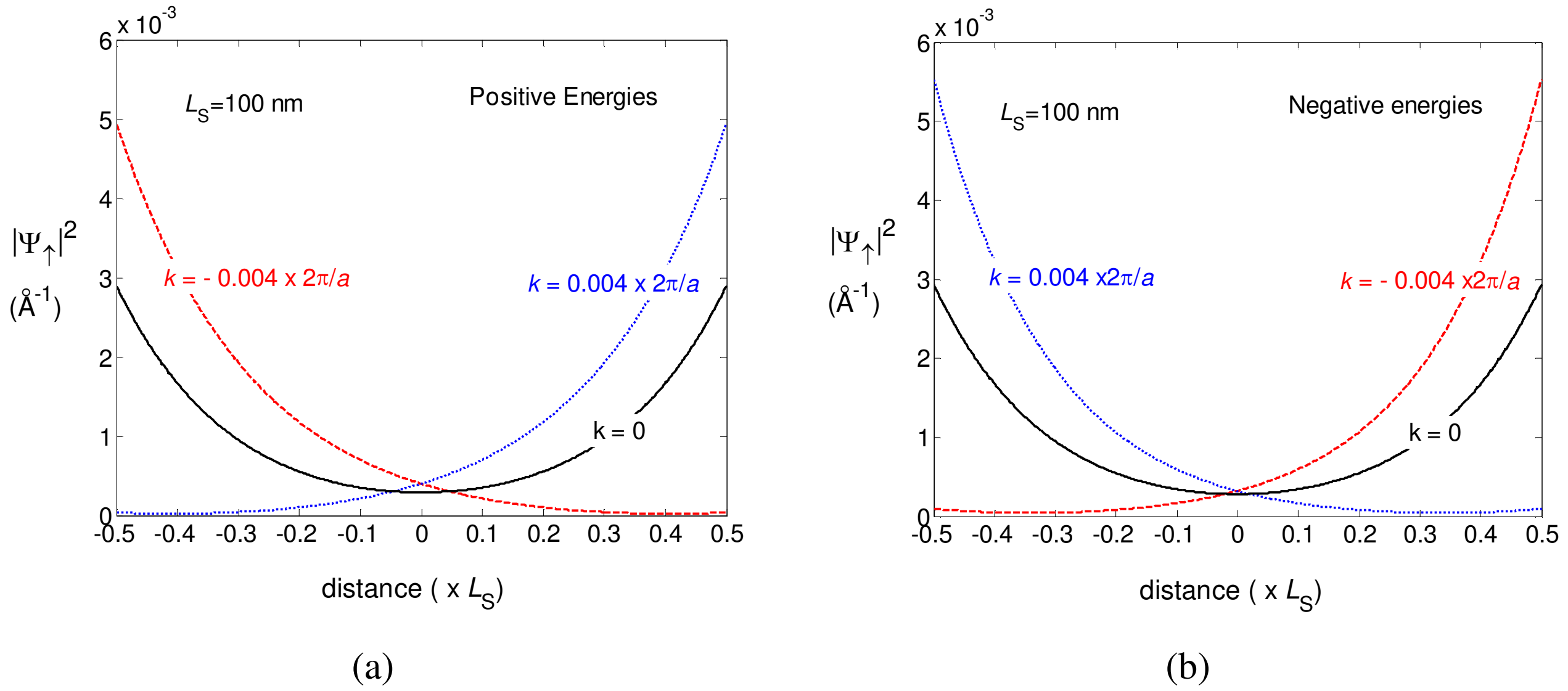

**Figure 5.** Probability densities corresponding to edge state wave vectors, $k$ = 0, and $k = \pm 0.004 \times 2\pi/a$, for (a) positive and (b) negative energy dispersions in Figure 4, where $a = a_{sub}$ is the cubic lattice parameter of the substrate (6.453Å). The edge on which the wave function is localized reverses, when either the sign of the wave vector or the energy is reversed.

The edge state probability density, $\left|\Psi_{\uparrow}\right|^2 = \left|\Psi_{1\uparrow}\right|^2 + \left|\Psi_{2\uparrow}\right|^2$ is calculated from Eq. (12), and is plotted at specific wave vectors for positive energies above the edge state band gap in Figure 5 (a), and for negative energies below the edge state band gap in Figure 5 (b). At zero wave vector, the wave function has equal probabilities on both edges, and a very low probability in between. Although the probability densities look similar, the wave function parameter $C_2$ in Eq. (12) is zero for the upper band edge, and has an infinite limit as the wave vector approaches zero for the lower band edge. At finite wave vectors the wave function becomes localized to one or the other edge, tending to the behavior of a wide sample. For example at a positive wave vector of $0.004 \times 2\pi/a$ it is localized to the right hand edge at positive energy and to the left hand edge at negative energy. When the wave vector is reversed, the sides also reverse ($a$ = 6.453 Å is the cubic lattice parameter of the substrate [11]).

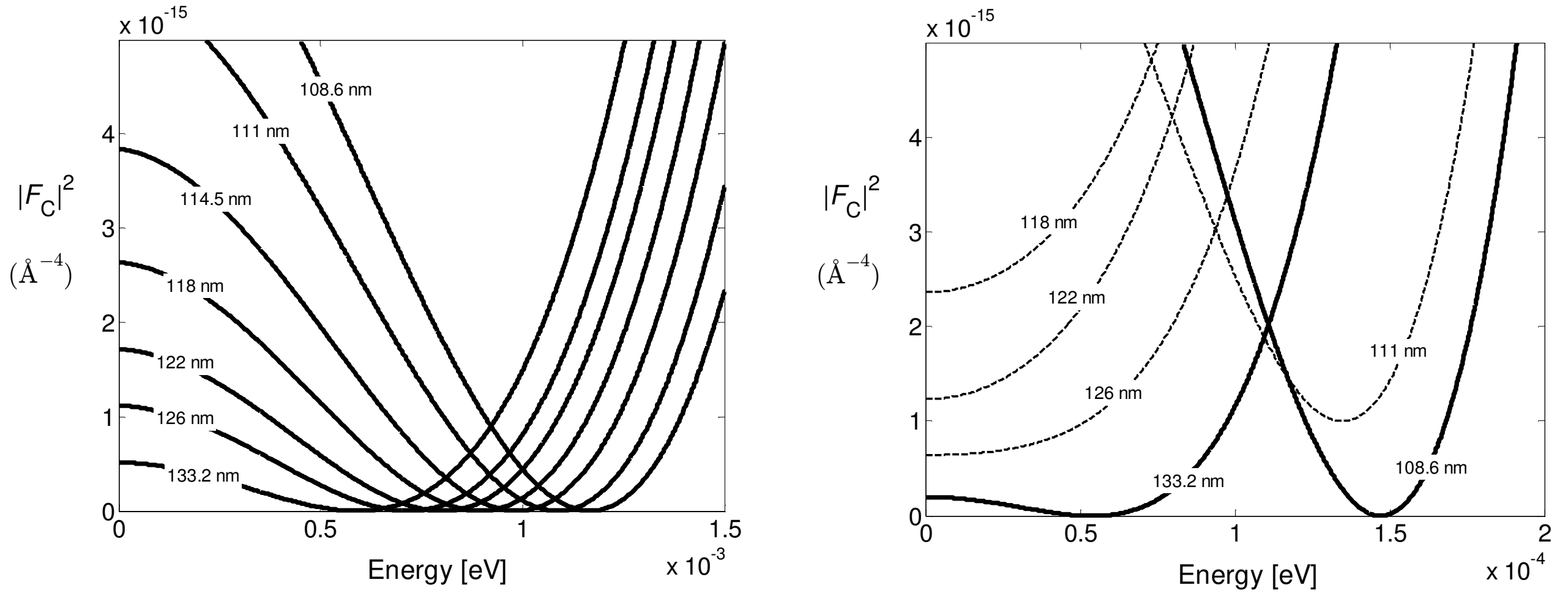

(a) (b)

**Figure 6.** Solution of the spin up characteristic equation, $|F_C|^2 = 0$ at $k$ = 0 as a function of sample width between 108 nm and 134 nm, for (a) a HgTe/CdTe QW, where the curves correspond to the non-spurious $\sigma_+$ solution in Eq. (4) and parameter set (2), and (b) an InAs/GaSb/AlSb QW where the curves are degenerate in $\sigma_+$ and $\sigma_-$, and parameter set (3) is used. In (b) solutions only exist at certain characteristic widths of 108.6 nm and 133.2 nm.

Figure 6(a) shows a plot of $|F_C|^2$ vs. positive energy at $k$ = 0 and different HgTe/CdTe sample widths. The energy at which $|F_C|^2 = 0$ is the solution to the characteristic equation, Eq. (15), and in this case yields the energy of the positive edge of the edge state band gap demonstrated in Figure 4. It can be seen in Figure 6(a) that this energy solution is a continuous function of the sample width, so the edge state band gap decreases monotonically with increasing width. Quite different behavior is observed in Figure 6(b) for a weakly hybridized InAs/GaSb/AlSb QW, using parameter set (3) in Table I. At this stage the simpler case is considered of a symmetric band structure with $D$ = 0. The asymmetric case with finite $D$ will be discussed later on. The function, $|F_C|^2$ in Figure 6(b) is identical for each complex conjugate decay parameter, $\sigma_\pm$ in Eq. (4), so when $|F_C|^2 = 0$ there are two degenerate solutions of the characteristic equation. Similar to HgTe/CdTe, a gap opens in the edge state spectrum at $k$ = 0 when the sample is narrow. However, Figure 6 (b) shows that there are positive energy solutions, $E_n$, only at specific sample widths, in contrast to the monotonic dependence on width shown in Figure 6(a) for the strongly hybridized case. A band gap, $2E_n$, thus appears in the edge state dispersions at certain "characteristic widths" which are given in nanometers by the formula $L_S$ ($n$) = 35 + 24.5$n$, where $n$ + 1 is the characteristic width index. Figure 6(b) therefore corresponds to the 4$^{th}$ and 5$^{th}$ characteristic width values, close to 108.5 and 133 nm, respectively.

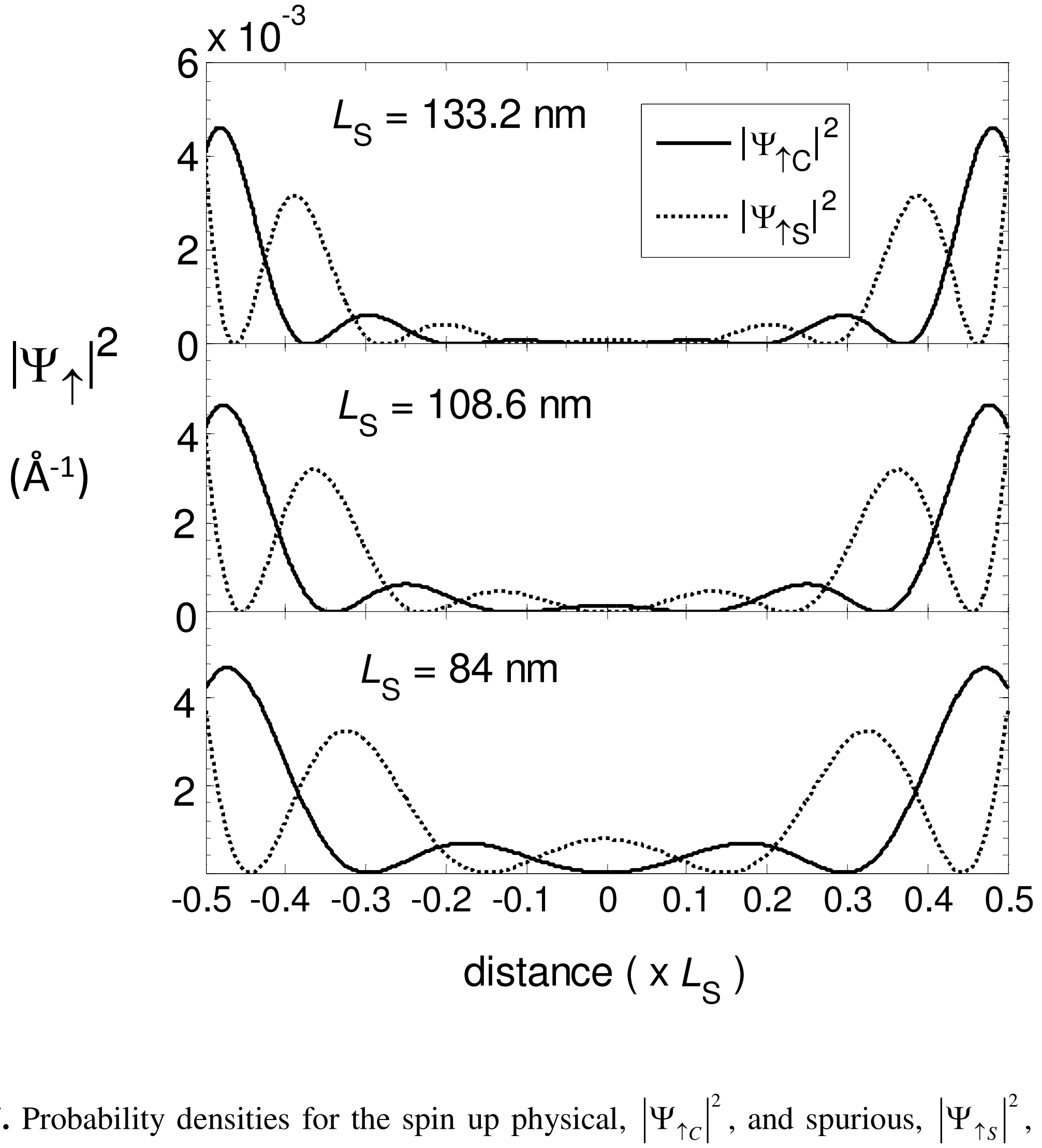

**Figure 7.** Probability densities for the spin up physical, $|\Psi_{\uparrow C}|^2$, and spurious, $|\Psi_{\uparrow S}|^2$, edge states at $k$ = 0 (solid and dashed lines, respectively), at characteristic widths of 84 nm, 108.6 nm and 133.2 nm (spin-down results are identical).

The reason for the existence of solutions only at certain characteristic widths of the weakly hybridized system can be understood by inspection of the wave functions. Figure 7 compares the probability densities at $k = 0$ for three adjacent characteristic widths of 84 nm, 108.6 nm and 133.2 nm. The solid lines are the physical solutions, $\left|\Psi_{\uparrow_C}\right|^2 = \left|\Psi_{1\uparrow_C}\right|^2 + \left|\Psi_{2\uparrow_C}\right|^2$, and the dashed lines are the spurious solutions, $\left|\Psi_{\uparrow_S}\right|^2 = \left|\Psi_{1\uparrow_S}\right|^2 + \left|\Psi_{2\uparrow_S}\right|^2$, based on linear combinations of the degenerate $\sigma_+$ and $\sigma_-$ solutions, as described in Ref. [12]. In the present case, this combination takes the form:

$$\overline{\Psi}_{\uparrow_{C,S}}\left(y\right)e^{ikx} = C_{\pm}\left\{\overline{\Psi}_{\uparrow,\sigma_-}\left(y\right)e^{-i\frac{\theta}{2}} \pm \overline{\Psi}_{\uparrow,\sigma_+}\left(y\right)e^{i\frac{\theta}{2}}\right\}e^{ikx} \qquad (19)$$

over the whole range of wave vectors, $k$, for which the decay parameters are complex, and where the positive sign is chosen for $\overline{\Psi}_{\uparrow_C}$, and the negative for $\overline{\Psi}_{\uparrow_S}$. Both solutions are real functions of $y$, so the probability current perpendicular to the edge is zero [27]. $C_{\pm}$ is a normalization constant and the phase factor, $\theta$, is adjusted to ensure orthogonality, noting that an increase or decrease by $\pi$ will switch $\overline{\Psi}_{\uparrow_C}$ to $\overline{\Psi}_{\uparrow_S}$ and vice versa [12]. It can be seen that the characteristic widths for $k = 0$ in Figure 7 are determined by the wave function oscillation period, with the probability density at the center of the sample changing from a node to an antinode between successive width values. For other sample widths, the absence of a solution to the characteristic equation suggests that the edge states are no longer purely exponential.

The edge state band gaps and characteristic widths can be reproduced quite well analytically, using Eq. (17). When $\sigma L > 1$, Eq. (17) (a) tends to the standard result for the decay parameters in very wide samples, given at the end of section IV. Thus, for strongly hybridized samples under these conditions, $\sigma\left(k=0\right) = \frac{A}{2B} - \sqrt{\frac{A^2}{4B^2} - \frac{|M|}{B}}$ can be substituted into Eq. (17) (b), yielding values for the edge state band gap, $\Delta E$, which agree to within 2% of those determined directly from the characteristic equation for $L_S \geq 100$ nm. In the case of weak hybridization, the decay parameters become complex [11], such that $\sigma(k=0)=\mu \pm i\upsilon$, approximating to $\mu=\frac{A}{2B}$ and $\upsilon=\sqrt{\frac{|M|}{B}-\frac{A^2}{4B^2}}$ when $\mathrm{Re}(\sigma L)>1$. The condition: Im ($\Delta E$) = 0, yields a relation for the characteristic widths:

$$\tan(\upsilon L_S)=\frac{\upsilon}{\mu} \tag{20}$$

with solutions $L_S(n')=9.9+24.7n'$ nm where $n'$ = 0, 1, 2,..., ∞, calculated using parameter set (3) in Table I. The energy gaps at these characteristic widths are given by:

$$\Delta E = 8|M|e^{-\mu L_S}|\cos(\upsilon L_S)| \tag{21}$$

which agrees to within a few percent with those deduced directly from the characteristic equation for $L_S \geq 84$ nm. The first characteristic width of 9.9 nm calculated from Eq. (20) is very narrow, yielding an energy gap of 0.0156 eV from Eq. (21) which is larger than the bulk band gap of 0.0116 eV at $k \approx 0.012\times 2\pi/a$ (see Figure 9 below). Ignoring this width value, the next characteristic width is 34.6 nm. The analytical characteristic widths, $L_S(n')$, then agree quite well with the values, $L_S(n)$, deduced above by solving the characteristic equation directly.

For a symmetrical band structure ($D$ = 0), it can be shown that the zone center energies at the edge state band gap are equally disposed about $E$ = 0, with values ± $\Delta E/2$, and the wave functions are $\begin{bmatrix}\phi_{1\uparrow}\\ \phi_{2\uparrow}\end{bmatrix}$ and $\begin{bmatrix}\phi_{2\uparrow}\\ \phi_{1\uparrow}\end{bmatrix}$. It has been shown in this section that $\phi_{1\uparrow}$, $\phi_{2\uparrow}$ are exponential (i.e. hyperbolic) functions always when electron-hole hybridization is strong, but only at the characteristic widths when hybridization is weak. For a semi-infinite sample the band gap closes and the zone center states are again exponential functions with zero energy. Therefore for widths between the characteristic values the band gap can never close because the non-exponential wave functions at these widths are deformed from the zero energy semi-infinite case, so they must have finite energies and thus a finite band gap. This is contrary to a recent numerical study which reported a periodic closing of the edge state bandgap as a function of sample width in weakly hybridized InAs/GaSb heterostructures [17]. If OBCs were used, which is suggested by the appearance of the wave function in this study, the vanishing of the band gap is perhaps another example of non-physical behavior that can occur for these boundary conditions. For weak hybrization, the $k$ = 0 OBC wave function of a semi-infinite sample with any value of $D$ varies as: $e^{-\mu' y}\sin\left(\upsilon' y\right)$, where $\sigma_{\pm} = \mu' \pm i\upsilon'$. This wave function has zero amplitude at the sample edge and also for successive nodes at periodic distances from the edge (see for example, the dotted curve in Fig. 3(b) of Ref. [12]). If the position of one of these nodes matches the width, $L_S$, of a narrow sample, this same wave function also satisties OBCs for the narrow case. Thus the energies are the same for the wide and narrow cases, and the band gap vanishes.

Returning to the SBC treatment, the wave vector dependence of the characteristic widths is now discussed, for which exponential solutions exist in weakly hybridized QWs. As shown previously by the author [12], the period of oscillation of the edge state wave function increases with edge state wave vector. An increasing period will cause the characteristic width to increase. Alternatively, for a fixed sample width, there are only solutions at certain characteristic wave vectors. Figure 8(a) shows the minimum value of $\left|F_C\right|^2$ as a function of wave vector for a constant sample width of 108.6 nm. $\left|F_C\right|^2$ oscillates and only satisfies the characteristic equation, $\left|F_C\right|^2 = 0$, at certain values of the wave vector. For other wave vectors, the characteristic equation appears to have no solution. Since there must be a solution that evolves smoothly with wave vector, the wave function must deform to accommodate the changing oscillation period, so that it has an envelope that is no longer purely exponential in character. Figure 8 (b) shows the probability density of the edge state wave functions based on Eq. (19) at wave vectors of $\pm 0.005775\,\frac{2\pi}{a}$ and an energy of +0.002708 eV, corresponding to the first point away from the origin at which $\left|F_C\right|^2 = 0$ in Figure 8 (a). It can be seen that the state which propagates in the positive $x$-direction is localized to the right hand wall, while the state propagating in the opposite direction is localized to the left hand wall, similar to the behavior shown for HgTe/CdTe at positive energies in Figure 5 (a). As for HgTe/CdTe, the edge states with a given wave vector switch sides when the energy is negative.

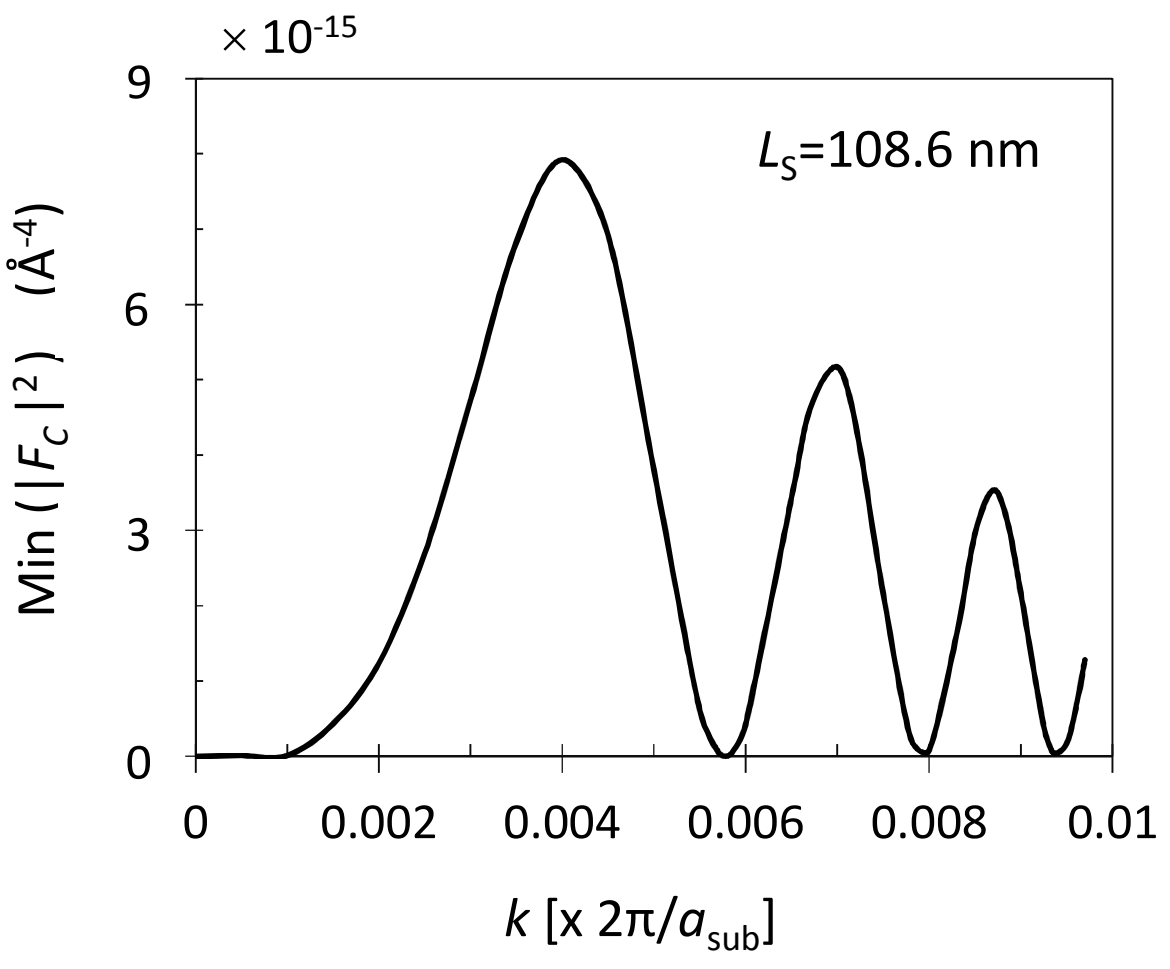

(a)

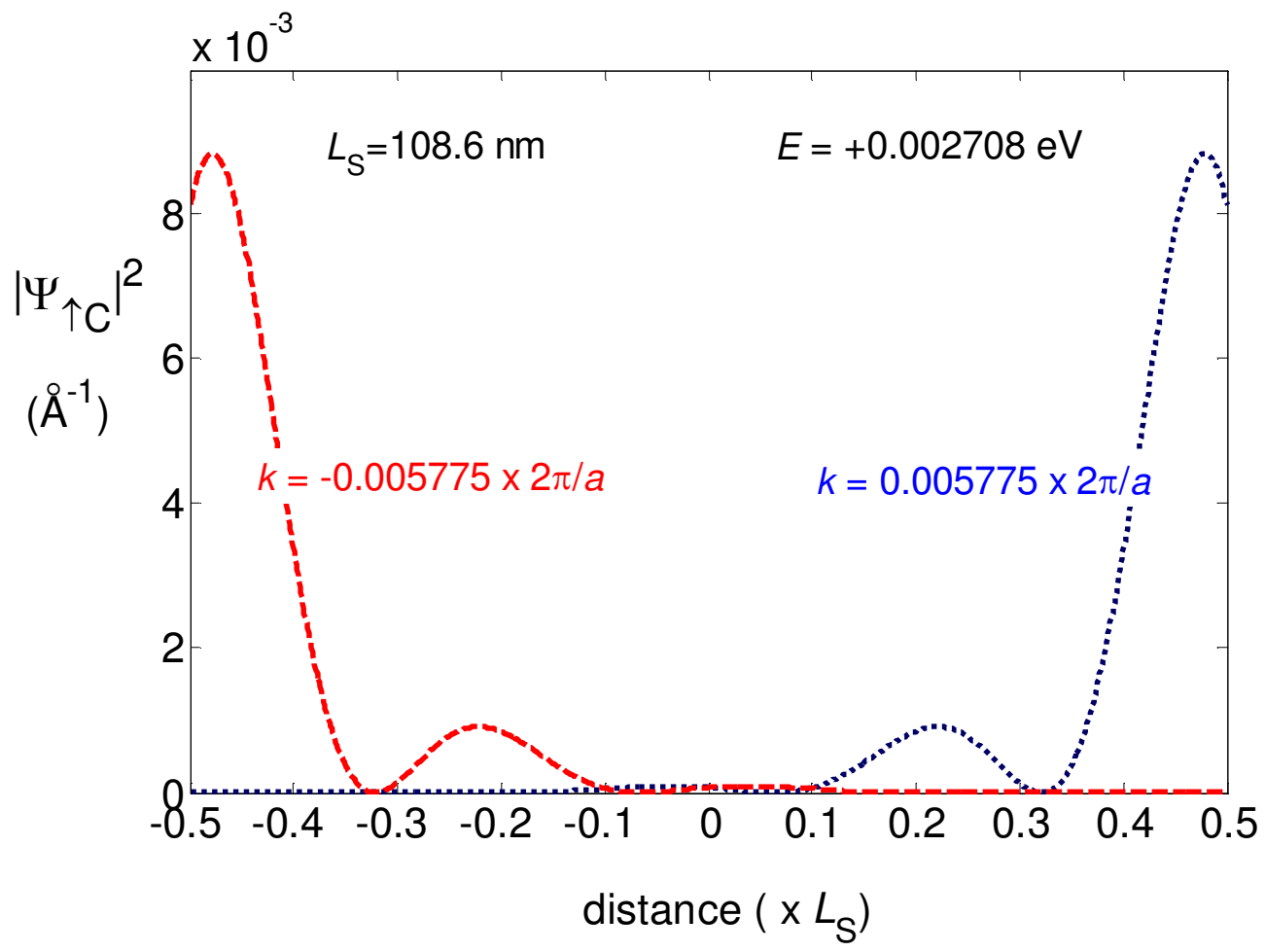

(b)

**Figure 8.** (a) Minimum value of $|F_C(\sigma_\pm)|^2$ vs. edge state wave vector for a InAs/GaSb/AlSb QW of sample width, $L_S$ =108.6 nm, using $A_0$ = $A$ and parameter set (3). Solutions of the characteristic equation, $|F_C(\sigma_\pm)|^2=0$, only exist at certain values of the wave vector. The curve applies to both complex conjugate values of the decay parameter, $\sigma_\pm$ , for which the solutions are degenerate, and does not depend on the signs of the wave vector or energy, or on the value of $A_0$ (b) Probability densities for the spin up physical edge states, $\left|\Psi_{\uparrow C}\right|^2$ , at $k = \pm 0.005775 \times 2\pi\, /a$ and $E$ = 0.002708 eV, corresponding to the shortest finite wave vectors for which $|F_C(\sigma_\pm)|^2=0$ in (a), when the solution has a purely exponential envelope function. The cubic lattice parameter of the substrate is $a = a_{sub}$ (6.0954 Å)**.**

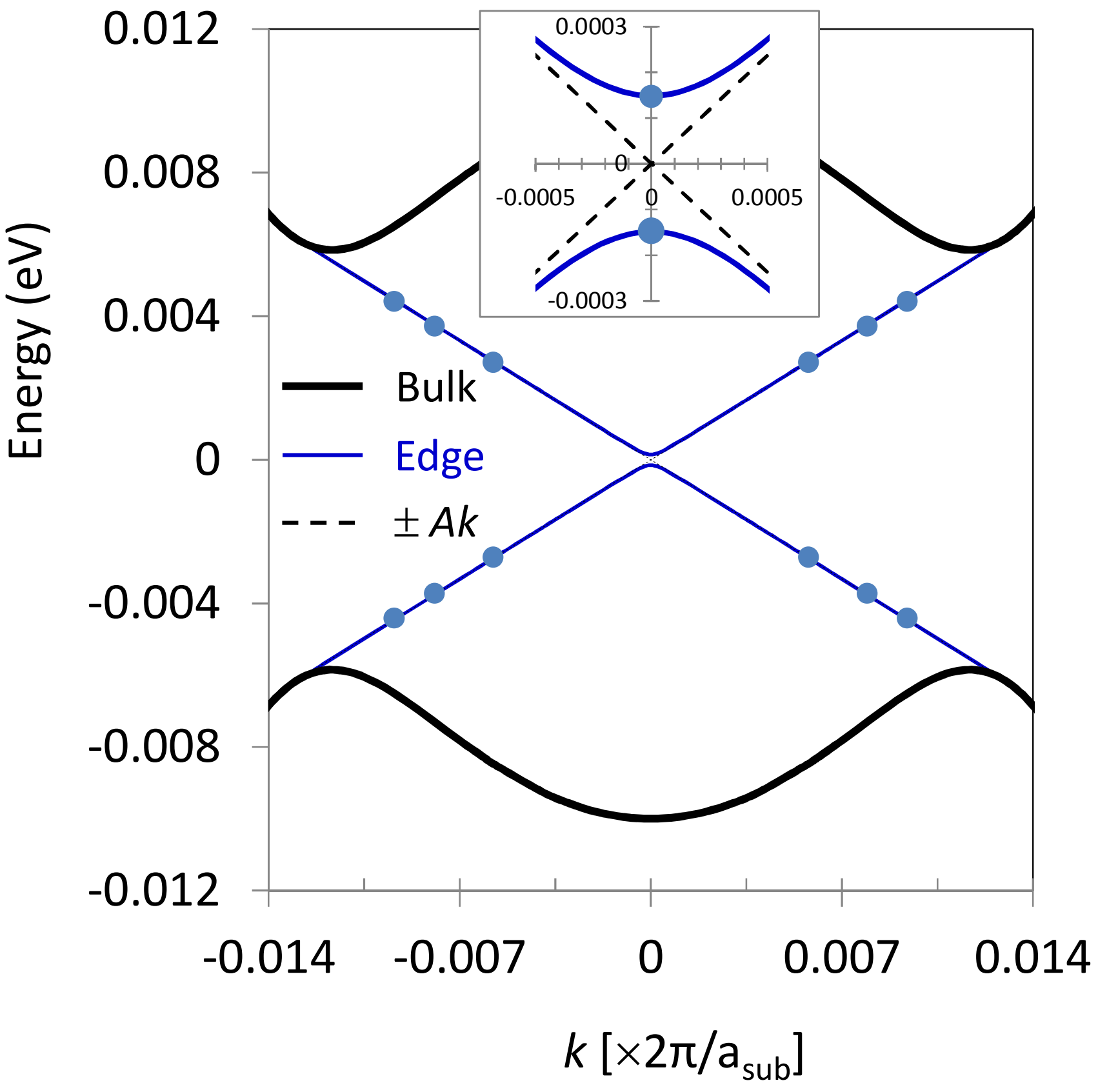

**Figure 9.** The bulk states (black lines) and spin up edge dispersions (blue lines) for the InAs/GaSb/AlSb QW of Figure 8. The edge dispersions are calculated from Eqs. (18) and (21) and are compared with edge dispersions, $\pm Ak$, for very wide samples (dashed lines). The points are calculated from the characteristic equation, $|F_C(\sigma_\pm)|^2=0$, which is satisfied at the wave vectors of the minima in Figure 8 (a). For clarity, the points at the zone center are only plotted in the inset, which presents this region on an expanded scale.

The full edge state dispersions of the weakly hybridized QW can be estimated from Eqs. (18) and (21), if it is assumed that the energies of the solutions which are not purely exponential can be interpolated accurately from those that are. The spin-up dispersions calculated in this way for the narrow sample of Figure 8 are presented as solid blue lines in Figure 9, where they are compared with the dispersions of the edge states in very wide samples, namely $\pm Ak$, which are plotted as dashed lines. The energies deduced from the characteristic equation, $|F_C|^2=0$, are also plotted as points, for the wave vectors where solutions exist. The agreement between the points and the full dispersions is very good, with a gap at the zone center of only 0.30 meV (i.e. twice the energy value for the 108.6 nm curve in Figure 6 (b)). This is an order of magnitude smaller than the gap found for the strongly hybridized QW in Figure 4, whose width is quite similar.

Figure 10 summarizes the dependence of the edge state band gap on sample width for both strongly and weakly hybridized systems. The weakly hybridized system is plotted as equally spaced points, calculated at the characteristic widths where an exponential solution exists at $k = 0$. The dashed line is simply an interpolation that has been assumed for the other sample widths with non-exponential solutions. As discussed above, the band gap should not vanish at widths between the characteristic values. For the strongly hybridized system, results of calculations are shown using parameter set (1) in Table I, so that a comparison can be made between the present SBC treatment and the OBC treatment proposed by Zhou et al. [13]. It can be seen that the SBC band gap is about a factor of two smaller than the OBC band gap for a sample width of 200 nm, and that this factor grows larger as the sample width increases. SBC results are also shown for parameter set (2), which was used in an earlier work by the author, as mentioned above. The SBC results using this parameter set are quite close to those calculated using parameter set (1), demonstrating only a weak sensitivity to the precise parameter values, and in particular, to whether $D > B$ as in parameter set (2), or $D < B$, as in parameter set (1). It should also be noted that the edge state band gap in the strongly hybridized system is several orders of magnitude larger than that in the weakly hybridized system, for sample widths greater than 200 nm.

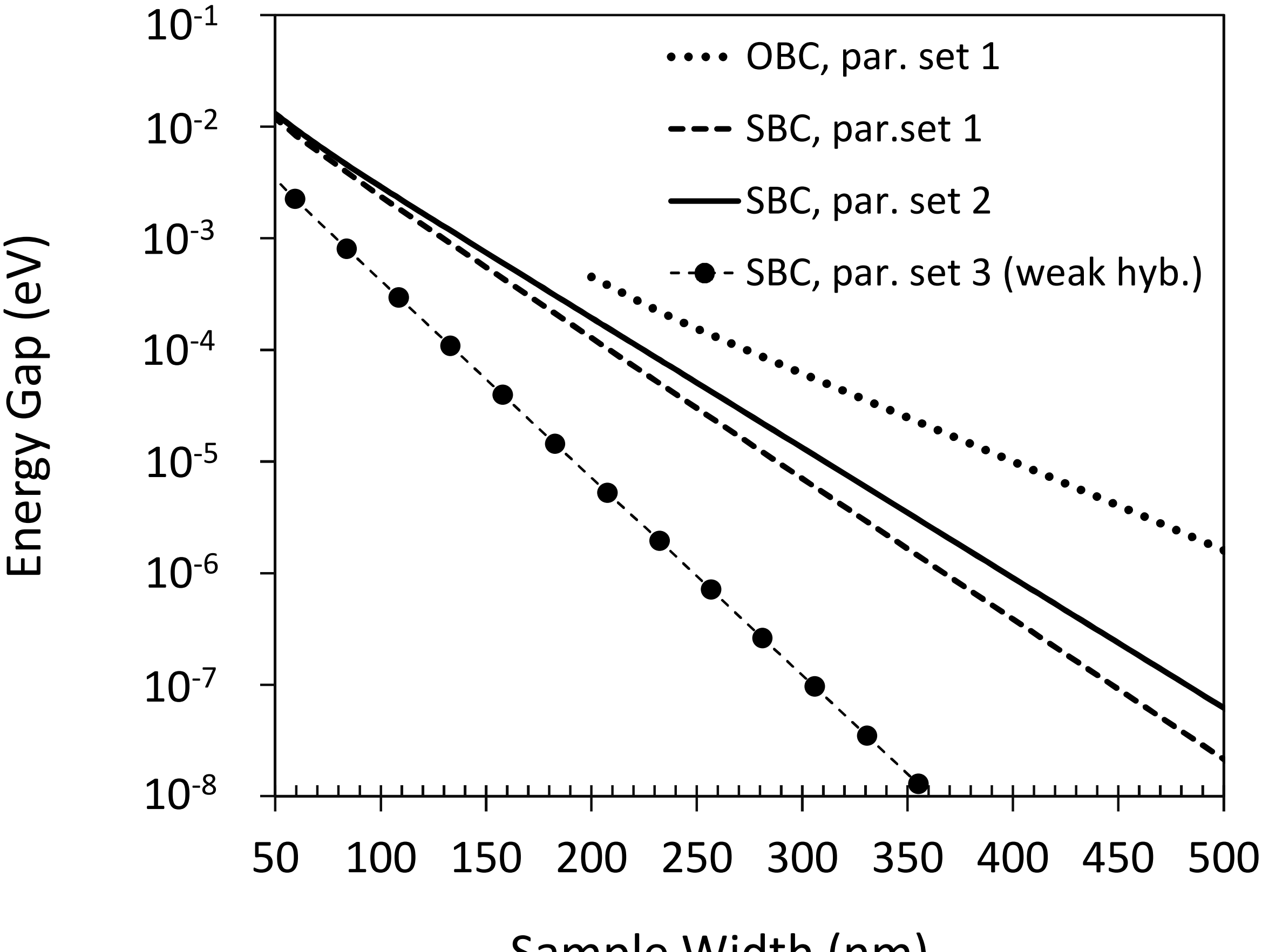

**Figure 10.** Dependence of the edge state band gap on sample width calculated with SBCs using parameter sets (1), (2) and (3). The results for parameter set (1) are compared with the OBC solution, reported for the same parameter values in Ref. [13].

The results presented in Figure 10 are essentially independent of the hybridization parameter, $A_0$, in the wall region. Eq. (13) shows that when $D$ is zero, the function $Q_D$ is one, and the characteristic equation in Eq. (15) becomes independent of $A_0$. Thus the results in Figure 10 for the weakly hybridized case with $D = 0$ do not depend on the wall hybridization parameter at all. For the strongly hybridized case at $k = 0$, using parameter set (2) and $A_0 = A$ yields $Q_D = 0.95$, and the value of $Q_D$ increases towards unity with increasing $A_0$. Thus even in the strongly hybridized case with a large $D$-value, the band gap results in Figure 10 are virtually independent of the wall hybridization parameter. The function, $Q_D$ also tends towards unity with increasing $k$, so the dispersion results presented for HgTe/CdTe in Figure 4 exhibit negligible dependence on $A_0$. When $A_0 = \hbar c >> A$, corresponding to the value for a free electron, the mid gap energy shifts by only -0.00050 eV relative to that shown in Figure 4 for $A_0 = A$.

The solid and dashed curves in Figure 10 highlight the fact that when electron-hole hybridization is strong, exponentially decaying edge solutions satisfy SBCs for any non-zero value of the band asymmetry parameter, *D*. However, when the electron-hole hybridization is weak this is not the case. Exponentially decaying edge solutions exist when $D = 0$ for wide samples, and for narrow samples with characteristic width values, but there are generally no purely exponential solutions over most of the wave vector range in weakly hybridized samples of any sample width when *D* is finite. Although the author reported edge state dispersion results for a semi-infinite InAs/GaSb/AlSb sample with *D* in the range 7-15 eV Å$^2$ [11, 12], these results turn out to be approximate, since they were based on finding the minimum in the value in $|F_E|^2$ in Eq. (16). Whilst this minimum is zero in the regions close to the bulk band edges where the edge state decay parameters are real (and the spurious solution splits off from the physical solution), closer to the center of the Brillouin zone the decay parameters in Eq. (4) become complex and the minimum in $|F_E|^2$, although very small, no longer goes precisely to zero. This observation, overlooked in previous work, can be taken as an indication that the edge state wave functions in this wave vector range are not purely exponential.

It can be argued, that the energy at which $|F_E|^2$ is a minimum is still a good approximation for wide weakly hybridized samples, because it occurs at essentially the same value at which $|F_E|^2 = 0$ if a small imaginary component is added to the edge state wave vector (with opposite signs for each of the two complex conjugate decay parameters). Clearly, the sample edge has translational symmetry in the direction of propagation, so a complex wave vector is not allowed. However, solutions of this sort should provide a good approximation to the exact solution in samples that are much shorter than the reciprocal of the imaginary wave vector component. It can easily be shown that this corresponds to sample lengths of at least several hundred Angstroms. Since the exact, non-exponential, solution must be independent of sample length, the short sample approximation must be close to the exact solution. In wide samples, the dispersions found for both spin directions using this approximation are not purely linear [12], and when $A_0 \approx A$, they have essentially the same $k$-dependence as for strongly hybridized systems with any value of $A_0 \geq A$, namely $Dk^2 \pm Ak$.

## VI. CONCLUSION

In spite of its simplicity, the 4 × 4 **k·p** Hamiltonian in a basis of electron and heavy hole ground states has been very successful in identifying many of the bulk and edge phenomena of two dimensional topological insulators. However, while some of its solutions describe actual physical states, others are spurious and exist in order to allow matching of boundary conditions at edges and interfaces and should not be interpreted as physically real. OBCs are attractive because they require zero amplitude at the sample boundaries and therefore avoid the need to define a Hamiltonian in the wall region beyond the boundaries. Unfortunately, edge states that satisfy OBCs incorporate both the real and spurious solutions, and while mathematically correct, they can exhibit unphysical characteristics. For this reason, it is impossible to avoid taking the wall region into account. Therefore, an approach has been followed where SBCs for the wave function and its derivative are obtained by integrating the Hamiltonian across the wall/semiconductor boundary. Exponentially decaying edge state solutions are found in the limit of an infinite wall potential, when the wave function decay parameter in the wall also becomes infinite, so the wave function is located entirely within the semiconductor. Based on this SBC approach, solutions were previously demonstrated for the physical spin-polarized edge states of a two dimensional topological insulator with a semi-infinite sample width, the amplitude of which is large at the edge and decays with distance into the semiconductor. Since the boundary conditions exert an influence over the properties of the edge states, these properties are changed from those deduced using OBCs, for example introducing non-linearity into the edge state dispersions. An important distinction is that the OBC wave function does not make contact with the edge while the SBC wave function is in intimate contact. In the SBC treatment, this makes the insensitivity of the edge states to edge imperfections even more remarkable, and is surely important for any proper calculation of the effect of such imperfections, especially magnetic ones which might flip the spin. In the present work, the SBC treatment has been extended to the case of samples with narrow widths, so that edge states from opposite boundaries interact and a band gap appears in the edge state energy spectrum. It has been found that some distinct differences occur in the behavior of materials with strong or weak electron-hole hybridization.

HgTe/CdTe is strongly hybridized and has physical, spin-polarized, exponentially decaying edge states with a real decay parameter and a non-linear energy dispersion that evolve smoothly with sample width. As the sample width is reduced, a gap opens at the zone center of the edge state dispersion, behavior qualitatively similar to that reported using OBCs. However, the band gap calculated with SBCs is about a factor of two smaller than the OBC band gap for a sample width of 200 nm, and this factor increases with sample width. The SBC band gap is relatively insensitive to the precise band structure parameters, including whether the band structure asymmetry is strong, $D > B$, or weak, $D < B$, and whether the electron-hole hybridization parameter in the wall is equal to or different from that in the sample. For strong band structure asymmetry there is no OBC solution, because the spurious solution switches from evanescent to oscillatory.

InAs/GaSb/AlSb is a weakly hybridized material. When the band structure asymmetry parameter is set to zero ($D = 0$), it was previously shown for very wide samples that there are degenerate, exponentially decaying, spin-polarized edge solutions with complex conjugate decay parameters over a specific range of positive and negative edge state wave vectors. The two complex conjugate solutions are combined into orthogonal physical and spurious solutions, where the physical solution evolves smoothly with wave vector over the region where the decay parameter becomes real, while the spurious solution does not. This picture changes dramatically as the sample width is reduced. A band gap opens in the edge state dispersion, but there are only exponential solutions at the zone center for certain characteristic widths. Using typical band structure parameters, the characteristic width in nanometers was found to obey the relation, $35 + 24.5n$, where $n + 1$ is the width index. The equal separation of the characteristic widths is related to oscillations in the exponentially decaying edge state wave function. The probability density at the center of the sample changes from a node to an antinode for successive characteristic width values. Although there are no purely exponential solutions for other sample widths, the band gap does not vanish and its width dependence has been deduced by interpolation from its values at the characteristic widths. It is found to be several orders of magnitude smaller than for strongly hybridized HgTe/CdTe QWs when the sample width is greater than 200 nm. For finite band structure asymmetry ($D \neq 0$), there are no exponential edge state solutions for any sample width, except close to the merging points of the edge states with the bulk band edges. Nevertheless, since typical $D$-values are quite small, it is assumed that the band gaps deduced for symmetric bands are still representative.

In all cases, the opening of a band gap in the edge state spectrum of narrow samples with an inverted band ordering yields spin degenerate conduction and valence states in both the bulk and edge regions. The material is then topologically trivial and when the Fermi level is located in both the bulk and edge state band gaps the sample is fully insulating at sufficiently low temperatures. Thus in addition to using top and bottom gates to locate the bulk Fermi-level close to mid-gap [4], it is possible to conceive of an experiment to verify some of the predictions in this work using narrow samples, typically 50-100 nm wide and 1-2 μm long, where a metal/insulator gate also covers a section of both edges. The fabrication of such structures should be possible using state of the art electron or ion beam lithography [28]. By applying a suitable bias to each edge gate, the Fermi level can be located in the edge state band gap and quantized conduction along the sample edges will be suppressed. In HgTe/CdTe QWs, this band gap is several milli-electron volts wide representing a significant fraction of the bulk band gap, so suppression of edge conduction should occur at temperatures of a few Kelvin. Figure 4 shows that in the SBC picture the edge state band gap lies in the middle of the bulk HgTe/CdTe band gap, so the suppression of edge conduction will be symmetrical with edge gate bias, over the range where the edge Fermi level is not pinned at either bulk band edge. On the other hand, in the OBC picture, the center of the edge state band gap lies just below the bulk conduction band and far above the bulk valence band [13]. Therefore, for this case the suppression of edge conduction should be very asymmetrical with respect to gate bias. Measurements of the dependence of edge conductance on edge gate bias can thus establish if SBCs indeed provide a more accurate physical picture.

For wide samples, it may be possible to resolve the edge state dispersion using Angle Resolved Photo-Emission spectroscopy (ARPES) on the edge of a wide multi-QW HgTe/CdTe sample. This type of measurement could potentially distinguish between the fairly symmetric dispersion predicted in the SBC picture, and the very asymmetric dispersion predicted in the OBC picture, as shown in Figure 2. Considerable success has been achieved with ARPES measurements on the edges of other topological insulator systems, for example see Ref. [29].

As mentioned above, an important consequence of using SBCs is that they should predict stronger scattering from point edge imperfections, particularly if magnetic impurities are involved that flip the spin, because their wave functions have a large amplitude at the sample boundaries, while for OBCs, this amplitude is zero. Therefore, the edge state coherence length calculated with SBCs should be shorter than that calculated using OBCs. Depending on the sample, this could affect the results of coherence length measurements similar to those reported in Ref. [10].

In summary, the SBC treatment with exponentially localized wave functions provides an effective description of the spin polarized edge states in strongly hybridized QW samples of any width, even in the presence of large band structure asymmetry. On the other hand, for weakly hybridized QWs, it only works at certain characteristic sample widths and edge state wave vectors, when there is no band structure asymmetry. In the presence of band structure asymmetry, purely exponential solutions do not exist over most of the wave vector range. Thus, a more sophisticated numerical approach based on SBCs will be needed to find the non-exponential solutions in these cases. The present work highlights both the importance of using boundary conditions which avoid spurious edge state solutions, and also the more complex nature of weakly hybridized systems.

**ACKNOWLEDGEMENT:** The author is indebted to Professor N. H. Linder in the Department of Physics, Technion, Haifa, 32000, Israel, for helpful explanations and comments.